\def\teq#1{$\, #1\,$}
\documentclass[11pt,dvips]{article}

\usepackage{picinpar}
\usepackage{wrapfig}
\usepackage{floatflt}
\usepackage{psfig}
\makeatletter
\renewcommand{\section}{\@startsection{section}{1}{0in}
	{0.4\baselineskip}{0.1\baselineskip}{\Large\bf}}
\renewcommand{\subsection}{\@startsection{subsection}{2}{0in}
	{0.25\baselineskip}{-\baselineskip}{\large\bf}}
\renewcommand{\subsubsection}{\@startsection{subsubsection}{3}{0in}
	{0.1\baselineskip}{-\baselineskip}{\normalsize\bf}}

\makeatother
\newcommand{\vol}[2]{$\,$\rm #1\rm , #2.}     
\def\teq#1{$\, #1\,$}
{\catcode`\@=11                                                  
\gdef\SchlangeUnter#1#2{\lower2pt\vbox{\baselineskip 0pt\lineskip0pt    
\ialign{$\m@th#1\hfil##\hfil$\crcr#2\crcr\sim\crcr}}}}

\def\apj{Astrophys. J.}                         
\def\apjl{Astrophys. J.}                        
\def\mnras{M.N.R.A.S.}                          
\def\nat{Nature}                                
\def\prl{Phys. Rev. Lett.}                      
\def\ssr{Space Sci. Rev.}                       

\begin{document}

%
\thispagestyle{myheadings}
%
\markright{OG.2.3.03}
\begin{center}
%
%
{\LARGE \bf Acceleration at Relativistic Shocks\\[4pt]
            in Gamma-Ray Bursts}
\end{center}

\begin{center}
%
%
{\bf Matthew G. Baring$^{1,2}$\\}
{\it $^{1}$Laboratory for High Energy Astrophysics, NASA/GSFC, Greenbelt, MD 20771, USA\\
$^{2}$Universities Space Research Association}
\end{center}

\begin{center}
{\large \bf Abstract\\}
\end{center}
\vspace{-0.5ex}
%
%
Most recent extragalactic models of gamma-ray bursts consider the
expansion of a relativistic blast wave, emanating from a solar-mass
type progenitor, into the surrounding interstellar medium as the site
for their activity.  The popular perception is that the optical
afterglows result from the external shock interface, while the prompt
transient gamma-ray signal arises from multiple shocks internal to the
expansion.  This paper illustrates a number of acceleration properties
of relativistic and ultrarelativistic shocks that pertain to GRB
models, by way of a standard Monte Carlo simulation.  Computations of
the spectral shape, the range of spectral indices, and the energy gain
per shock crossing are presented, as functions of the shock speed and
the type of particle scattering.
%

\vspace{1ex}

\section{Introduction}
\label{intro.sec}
The gamma-ray burst (GRB) field has burgeoned in the last few years,
and particularly after the discovery (e.g. van Paradijs, et al. 1997;
Costa, et al. 1997; Frail, et al. 1997) of
optical, radio and X-ray transient counterparts, and the
subsequent identification of a cosmological redshift via atomic
absorption features in the 8th May 1997 afterglow (Metzger et al.
1997).  Theoretical interpretations abound, mostly focusing on some
variation of a blast wave expansion impacting on the interstellar
medium (ISM) surrounding the burst (e.g. Meszaros \& Rees 1993).
Supersonic blast wave impact upon the surrounding ISM
guarantees relativistic shock formation, basically a relativistic
version of supernova remnants; the dissipation of the ram pressure
kinetic energy via diffusive particle acceleration in the shock is the
commonly-invoked means of converting the bulk motion into a viable
supply of energy for radiative purposes.  Principal quantities
generally appearing in radiation emission models of GRBs include the
Lorentz factor \teq{\Gamma_1} of the shock, and the spectral index of
the electron (and perhaps ion) population.  Furthermore, modelling of
ultra-high energy cosmic ray production by GRBs (e.g. Waxman
1995; Vietri 1995) requires knowledge of the mean energy gain a
particle experiences in traversing the shock.  To date, GRB
models of both the transient gamma-ray event and fireball afterglows
have invoked only the crudest notions of Fermi acceleration.  The aim
of this presentation is to probe these shock acceleration properties,
which are pertinent to GRB blast wave models, using results from a
Monte Carlo simulation of diffusive acceleration.

The Monte Carlo technique we employ here has been described in detail
in numerous expositions (Ellison, Jones \& Eichler 1981; Jones and
Ellison 1991; Baring, Ellison \& Jones 1993; Ellison, Baring, \& Jones
1996).  The simulation technique is a kinematic model.  Particles are
injected upstream and allowed to convect into the shock, colliding with
postulated scattering centers (presumably magnetic irregularities)
along the way.  As they diffuse between the upstream and downstream
regions, they continually gain energy.  An important property of the
model is that it treats thermal particles like accelerated ones, making
no distinction between them. Hence, as the accelerated particles start
off as thermal ones, this technique automatically injects particles
from the thermal population into the acceleration process.  One
valuable consequence of this unified treatment is that modification of
the shock hydrodynamics by the accelerated population can easily be
incorporated.  Such non-linear hydrodynamics are omitted in the present
{\it test-particle} application, though they will probably be an
important aspect of the GRB acceleration problem given their relevance
to the modelling of supernova remnant emission (e.g. Baring et al.
1999).

The test-particle results presented here use a guiding-center version
of the Monte-Carlo technique, older than our latest codes which compute
particle gyro-orbits exactly rather than just track the center of
gyration.  The guiding-center approach, which is detailed in Baring,
Ellison \& Jones (1993), is often expedient, and is entirely
appropriate to plane-parallel shock applications (where the field lies
along the shock normal, i.e. \teq{\Theta_{\rm B1}=0}), which form the
focus here.  This method is precisely that implemented in Ellison,
Jones \& Reynolds' (1990, hereafter EJR90) treatment of parallel
relativistic shocks, thereby providing a principal motivation for
adhering to a similar approach.  The updated code replicates results
obtained in EJR90.  Following EJR90, both large angle scattering (LAS)
and pitch angle diffusion (PAD) will be implemented here.  
For LAS, the mean-free path \teq{\lambda} in the fluid frame is
constrained to be proportional to a particle's gyroradius \teq{r_g}, in
accord many previous expositions (e.g. EJR90, Baring, Ellison \& Jones
1993, hereafter BEJ93; Ellison, Baring, \& Jones 1996), while for PAD
we set \teq{\lambda} to be independent of \teq{r_g}, following EJR90.

\section{Results}
 \label{results.sec}
The primary purpose of this paper is to extend the work of EJR90 to
ultrarelativistic shocks, and to explore spectral properties of results
from our simulation in the context of gamma-ray bursts.  Representative
ion distributions obtained in the rest frame of the shock are depicted
in Figure~1 for \teq{\Gamma_1=5}.  The notations used are
\teq{\Gamma_1} (\teq{\Gamma_2}) for the Lorentz factor of the upstream
(downstream) flow speed \teq{u_1} (\teq{u_2}) in the shock rest frame,
and \teq{r} for the velocity compression ratio in this frame.
The value \teq{r=3} is chosen to mimic expectations from an
ultrarelativistic \teq{\gamma=4/3} gas, though other values are
possible due to possible mildly-relativistic nature of the downstream gas.
The upstream gas temperature and magnetic field are kept
sufficiently low to maintain the strength of high Mach number shocks
and reduce the number of parameters to which results are sensitive.

The left panel of Figure~1 illustrates how a smooth power-law spanning
many decades can be obtained at energies not too far in excess of
thermal ones.  The spectrum is somewhat steeper, though not markedly
so, than the canonical \teq{E^{-2}} particle distribution for
strong non-relativistic shocks.  For slower shock speeds, the
PAD-generated power-laws are steeper still (e.g. Kirk \& Schneider
1987; EJR90), while for faster shocks appropriate to GRB scenarios, the
spectral index \teq{\sigma} saturates around the value of 2.2 when
\teq{\Gamma_1\gg 1}, as found by Bednarz \& Ostrowski (1998).  Index
determinations are listed in Table~1, and while here we replicate
Bednarz \& Ostrowski's (1998) asymptotic behaviour, we also find that
\teq{\sigma} monotonically decreases with \teq{\Gamma_1}, in
contradiction to their findings in the range around \teq{\Gamma_1\sim
3}.  Since the simulation reproduces the analytic results of Kirk \&
Schneider (1987) at \teq{u_1=0.9c} of \teq{\sigma} considerably higher
than 2.2, our findings of monotonicity appear plausible.

\begin{center}
  \begin{tabular}{cccc}
     \hline\\[-10pt]
     & & PAD & LAS \\
     \teq{\Gamma_1} & \teq{u_1/c} & (\teq{\lambda\propto r_g^0}) 
               & (\teq{\lambda\propto r_g}) \\[3pt]
     \hline\\[-11pt]
     \hline\\[-11pt]
     2.29 & 0.9 & 2.34 & 1.81 \\
     3 & 0.9428 & 2.23 & 1.59 \\
     5 & 0.9798 & 2.22 & 1.49 \\
     9 & 0.9938 & 2.20 & 1.41 \\
     27 & 0.9993 & 2.19 & - \\
     81 & 0.9999 & 2.18 & - \\
     \hline
  \end{tabular}
\end{center}
Table 1: Asymptotic spectral indices \teq{\sigma} for plane-parallel
(\teq{\Theta_{\rm B1}=0^\circ}) relativistic shocks of various Lorentz
factors \teq{\Gamma_1} and compression ratio \teq{r=3} for the cases of
pitch angle diffusion (PAD), and large angle scattering (LAS).
Simulational uncertainties in predicting \teq{\sigma} are typically of
the order of 1--2\% for PAD, and twice that for LAS.  Spectra for the
\teq{\Gamma_1=5} case are exhibited in Figure~1.


The right panel of the Figure depicts a spectrum obtained for LAS with
similar parameters.  Two striking features emerge: (i) that the
asymptotic power-law is generated only at much higher energies than in
the PAD case, and (ii) when it is obtained, it is much flatter than for
PAD.  Both are essentially due to the prompt removal of particles by
single large angle scatterings from the narrow Lorentz cone of
directions for which the ions can remain upstream of the shock.   PAD
and LAS generate different angular distributions ahead of a
relativistic shock, and the fact that these result in dissimilar
spectral indices is widely understood (e.g. see EJR90 and references
therein).  The spectral structures above the thermal peak for LAS
correspond to contributions from successive shock crossings in a manner
somewhat like the structure seen at mildly suprathermal energies in
non-relativistic shocks (e.g.  BEJ93).  These structures start out flat
due to a kinematic spread induced by LAS, and then slowly steepen and merge
into the power-law continuum.  The trend for LAS is again that for
faster shocks, \teq{\sigma} declines (see Table~1).  However, any
putative saturation could not be demonstrated numerically, due to
difficulties in generating good statistics at energies well in excess
of \teq{10^{20}}eV.

\begin{figure}[t]
\centerline{\hskip 0.0truecm
             \psfig{figure=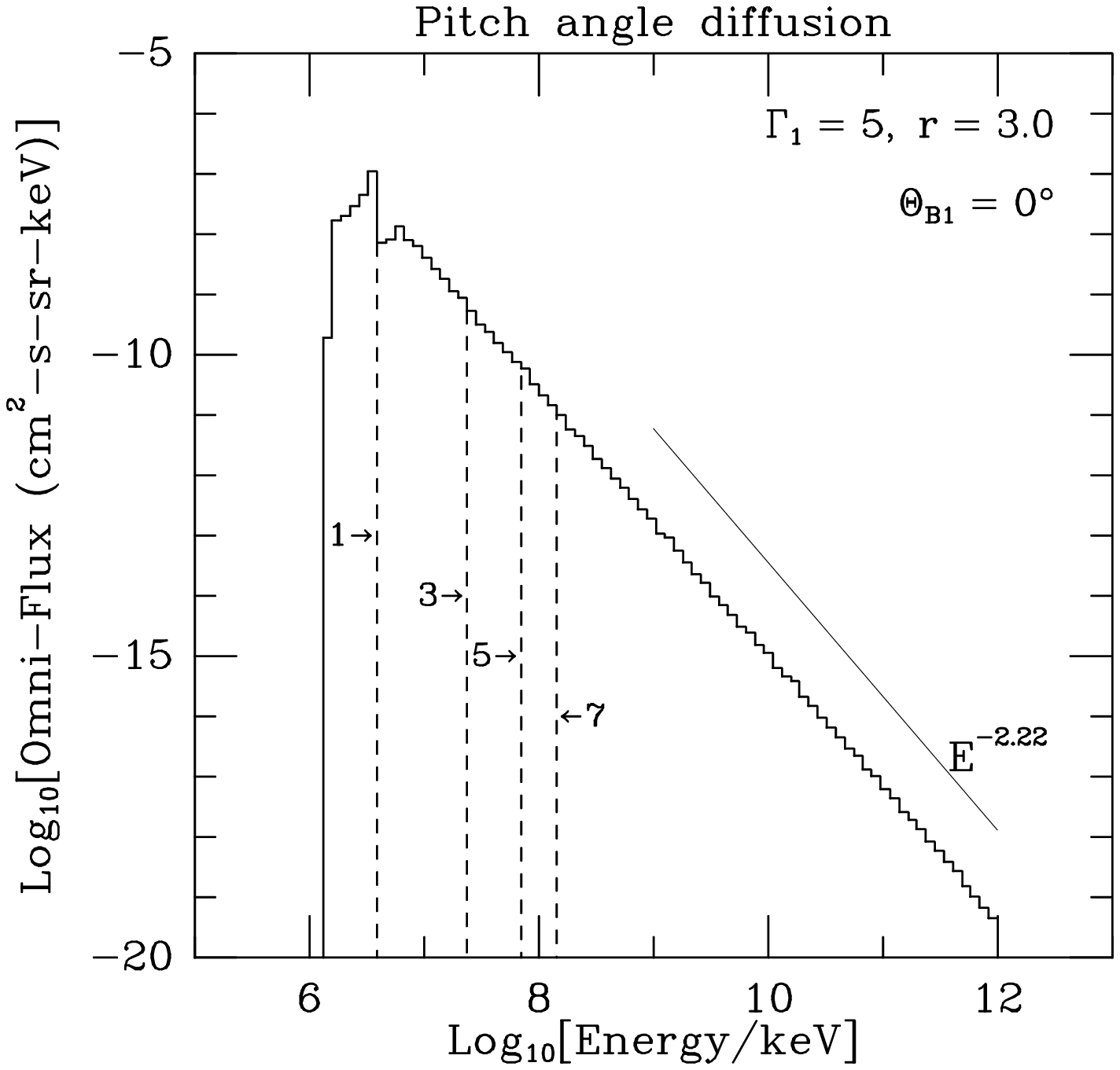,width=7.9cm}\hskip 0.5truecm
              \psfig{figure=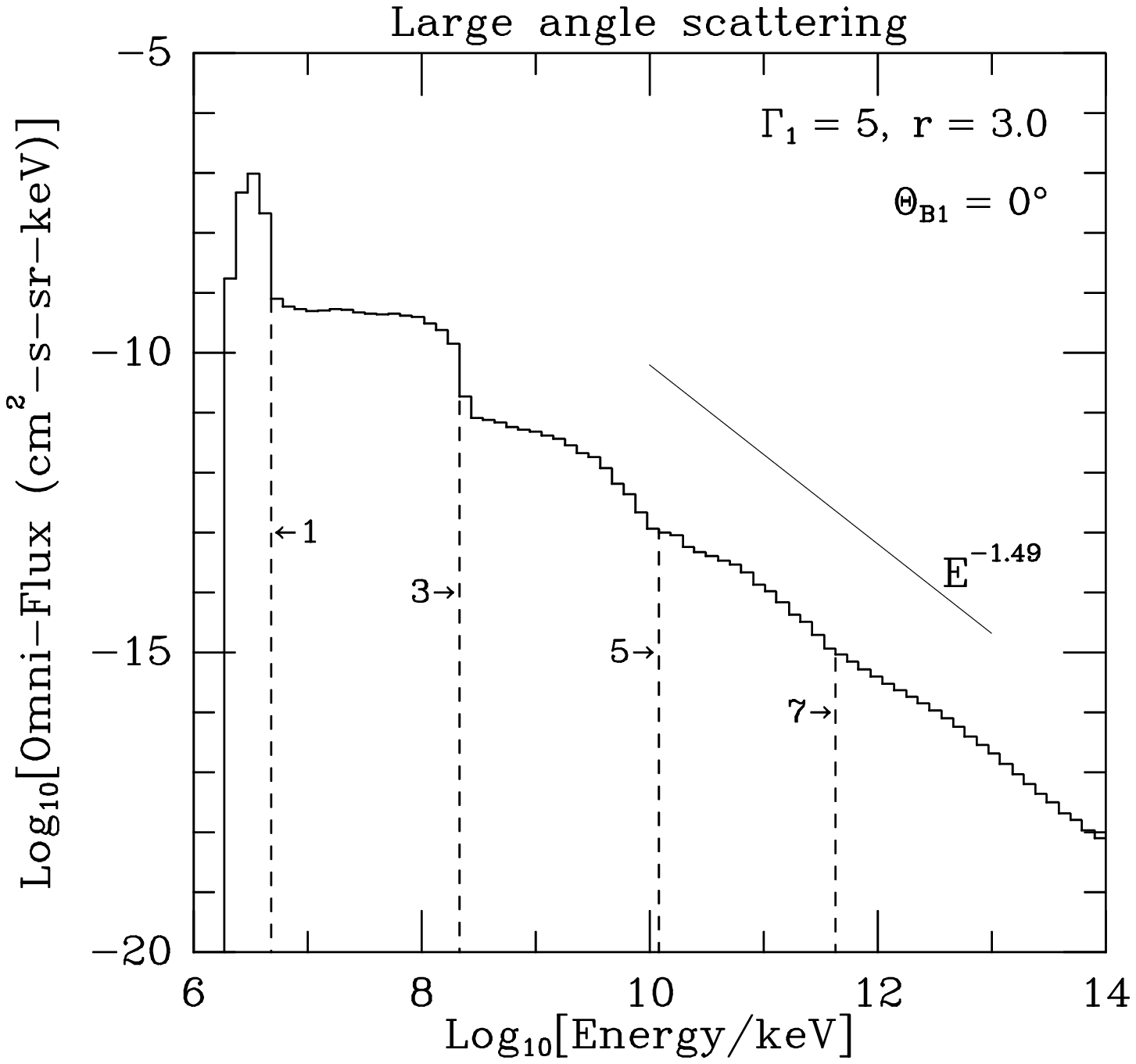,width=8.1cm}}
\caption{
Omni-directional fluxes for protons (of mass \teq{m_p}) accelerated in
relativistic shocks of Lorentz factor \teq{\Gamma_1} (i.e. upstream
flow speed \teq{c\, [1-1/\Gamma_1^2]^{1/2}} and velocity compression
ratio \teq{r=3} in the shock rest frame.  These high Mach number,
test-particle shocks are plane-parallel (\teq{\Theta_{\rm
B1}=0^\circ}).  The fluxes represent differential (in energy)
distributions multiplied by the particle speed (i.e.  \teq{\approx c}),
as measured in the shock frame somewhat downstream of the shock
discontinuities.  Cases corresponding to two types of scattering are
illustrated, namely pitch angle diffusion (PAD; Left Panel), and large
angle scattering (LAS; Right Panel).  In each case, in addition to the
total downstream spectrum, the maximum energies corresponding to 1, 3,
5, and 7 shock crossings are labelled, indicating the generally
monotonic increase in energy that is a signature of the Fermi
mechanism.  Observe that for PAD, this increase falls far short of an
amplification by a factor of \teq{\Gamma_1^2} at higher energies.  The
asymptotic power-law behaviours, which are achieved at energies
\teq{\gg \Gamma_1 m_pc^2}, are indicated.
}
   \label{fig:icrc99psr_f1}
\end{figure}

The maximum energy of particles in the downstream region for 1,3,5 and
7 shock-crossings are also depicted as vertical dashed lines in
Figure~1.  For LAS, it is clear that this scales as \teq{\Gamma_1^2},
and energy amplification per shock crossing that is widely (and
erroneously) quoted in GRB model literature.  This represents the
maximum amplification, and the {\it mean amplification is much less},
declining with increasing energy.
This contention follows immediately from the tendency of the
spectral plateaux to steepen with energy, given that the probability of
convection downstream of the shock drops with increasing energy.  For
PAD, even the maximum amplification falls far short of \teq{\Gamma_1^2},
and saturates to a factor of order unity at high energies.  This can be
seen as follows.  For a downstream particle of speed \teq{\beta c} (in
the shock frame) that crosses upstream, its velocity angle relative
\teq{\theta} to the shock normal must satisfy \teq{\beta\cos\theta
>\beta_2} (\teq{\sim 1/3}).  This yields a range of possible upstream
fluid-frame Lorentz factors \teq{\gamma_{\rm F}} given by
\teq{\gamma\Gamma_1 (1+\beta_1\beta_2) < \gamma_{\rm F} <
\gamma\Gamma_1 (1+\beta_1\beta)}, beamed in a narrow cone around
\teq{\cos\theta_{\rm F}=1}.  Pitch angle diffusion gradually widens
this distribution till \teq{1-\cos\theta_{\rm F}'\sim 1/(2\Gamma_1^2)},
at which point the ions convect downstream again with a shock frame
Lorentz factor \teq{\gamma'=\gamma_{\rm F}\Gamma_1 (1-\beta_{\rm
F}\beta_1\cos\theta_{\rm F}')}.  It is then trivial to determine that
\teq{\gamma'/\gamma\sim 2} (\teq{\ll\Gamma_1^2}), a result noted by
Gallant \& Achterberg (1999).

The implications of these results for gamma-ray burst modelers are the
following.  First, liberal use of \teq{\Gamma_1^2} amplification
factors in shock crossings when estimating maximum energies obtainable
in relativistic shocks is inappropriate.  This impacts contentions
(Waxman 1995; Vietri 1995) that GRBs can generate ultra-high energy
cosmic rays, as do reductions in acceleration times seen when
\teq{\Gamma_1\gg 1} (e.g. see EJR90, and references therein).  The
disparate nature of the spectral indices between PAD and LAS cases is
also of concern to the GRB community.  While PAD yields a narrow range
of indices more-or-less commensurate with inferences from both prompt
gamma-ray emission and delayed X-ray/optical afterglows, the
structured, flat LAS spectra may be at odds with GRB data.
Furthermore, given the broad dynamic range, huge differences would
arise in flux predictions for different wavebands.  While the results
presented here are for protons, one expects similar spectral properties
for (and hence the concerns for emission from) electrons, since they
too are relativistic when \teq{\Gamma_1\gg 1} and hence readily
resonate with Alfv\'en and whistler modes.  Bednarz \& Ostrowski (1996)
argue in favor of PAD operating in shocks with \teq{\Gamma_1\gg 1}.  We
believe the situation not to be transparent.  The definition of PAD in
the context of relativistic shocks is effectively that angular
deflections are substantially within the Lorentz loss-cone of
half-angle \teq{1/\Gamma_1}.  As soon as deflections exceed this small
value, spectra from our simulations quickly flatten to reproduce the
LAS ones exhibited here.  Hence the critical issue as to whether PAD or
LAS operates in GRBs is contingent upon the typical magnitude of
particle deflections in field turbulence associated with relativistic
shocks.  This nontrivial question will be the subject of future
investigation.


%
\vspace{-3pt}
\vspace{1ex}
\begin{center}
{\Large\bf References}
\end{center}
\vspace{-3pt}
Baring, M.~G., Ellison, D.~C. \& Jones, F.~C. 1993, \apj\vol{409}{327} (BEJ93)
\\
Baring, M.~G., Ellison, D.~C., Reynolds, S.~P., Grenier, I.,
 \& Goret P. 1999, \apj\vol{513}{311}
\\
 Bednarz, J. \& Ostrowski, M. 1996, \mnras\vol{283}{447}
\\
 Bednarz, J. \& Ostrowski, M. 1998, \prl\vol{80}{3911}
\\
 Costa, E., et al. 1997, \nat\vol{387}{783}
\\
 Ellison, D.~C., Baring, M.~G., \& Jones, F.~C. 1996, \apj\vol{473}{1029}
\\
 Ellison, D.~C., Jones, F. C., \& Eichler, D. 1981, J. Geophysics \vol{50}{110}
\\
 Ellison, D. C., Jones, F. C. and Reynolds, S.~P. 1990, \apj\vol{360}{702}
 (EJR90)
\\
 Frail, D.~A., et al. 1997, \nat\vol{389}{261} 
\\
 Gallant, Y.~A., \& Achterberg, A. 1999, \mnras\ in press.
\\
 Jones, F.~C., \& Ellison, D.~C. 1991, \ssr\vol{58}{259}
\\
 Kirk, J.~G. \& Schneider, P. 1987 \apj\vol{325}{415}.
\\
 M\'esz\'aros, P. \& Rees, M.~J. 1993, \apj\vol{405}{278}
\\
 Metzger, M.~R., et al. 1997, \nat\vol{387}{878} 
\\
 van Paradijs, J., et al. 1997, \nat\vol{386}{686}
\\
 Vietri, M. 1995, \apj\vol{453}{883}
\\
 Waxman, E. 1995, \apjl\vol{452}{L1}
\\

\end{document}